\documentclass[aps,prb,reprint,showpacs,citeautoscript]{revtex4-1}

\usepackage{graphicx}
\usepackage{dcolumn}
\usepackage{bm}
\usepackage{epsfig}
\usepackage{amsmath}
\usepackage[usenames,dvipsnames]{color}
\usepackage{booktabs}
\begin{document}

\author{Mahmoud M. Asmar}
\email{asmar@lsu.edu} \affiliation{Department of Physics and Astronomy, Louisiana State University, Baton Rouge, LA 70803-4001}
\author{Sergio E. Ulloa}
\email{ulloa@ohio.edu} \affiliation{Department of Physics and
Astronomy and Nanoscale and Quantum Phenomena Institute, Ohio
University, Athens, Ohio 45701-2979}

\title{Minimal Geometry for Valley Filtering in Graphene}

\begin{abstract}
The possibility to effect valley splitting of an electronic current in graphene represents the essential component in the new field of valleytronics in such two-dimensional materials. Based on a symmetry analysis of the scattering matrix, we show that if the spatial distribution of multiple potential scatterers breaks mirror symmetry about the axis of incoming electrons, then a splitting of the current between two valleys is observed. This leads to the appearance of the valley Hall effect. We illustrate the effect of mirror symmetry breaking in a minimal system of two symmetric impurities, demonstrating the splitting between valleys via the differential cross sections and non-vanishing skew parameter.  We further discuss the role that these effects may play in transport experiments.
\end{abstract}
\maketitle

\textit{Introduction}. The honeycomb arrangement of carbon atoms in graphene results in and electronic spectrum
characterized by linearly dispersing cones near the Fermi level at opposite corners of the Brillouin zone.
These regions or {\em valleys} host effectively massless electronic excitations characterized by their helicity.
This quantum number is associated with the intrinsic structure of the eigenstates, described in terms of the pseudo-spin
degree of freedom that identifies different atoms in the unit cell.

Although typical current injection in graphene populates both valleys equally, the nascent
field that aims to control, manipulate and utilize the valley degree of freedom in electronic applications is known as
{\em valleytronics} \cite{vallfilter,valleyline,valleyline2,valleyline3,valleyline4,valleymagnetic,valleymassbarrier,%
valleymassbarrier2,valleyphonon,valleytrigonal,valleystrain,valleystrain1,starinPeters,lopesvalleyhall}.
This term is in analogy to spintronics, which aims to access and control the spin degree of freedom in devices \cite{spintronics,sinova}.
As two well-defined valleys are seen not only in graphene but in other two dimensional materials, such as
transition metal dichalcogenides \cite{layered,layered1}, there is a great deal of interest in understanding the essential
elements and characteristics of valleytronic implementation and applications in these 2D Dirac materials \cite{DiracMatRev2014}.

Controlling the valley degree of freedom obviously requires the differentiation between degenerate valleys via controllable perturbations.
Several perturbations have been identified  to achieve such differentiation and allow the control and production of valley polarized currents.
Approaches include the use of magnetic fields and barriers \cite{valleymagnetic},  artificial or naturally occurring lattice
deformations, which result in valley-dependent pseudo-magnetic fields \cite{valleystrain,valleystrain1,ramon1,starinPeters},
and time-dependent lattice vibrations \cite{valleyphonon}.  Hybrid layered systems, such as graphene on hexagonal boron nitride ($h$BN) \cite{topologicalvalley, hallsttagered}, graphene separated by $h$BN \cite{chiralandvalley}, and graphene on transition metal dichalcogenides \cite{abdulrahman}, have been shown to allow control of the valley quantum number and successful generation of valley currents, via the relative orientation of the lattices, and/or in-plane magnetic fields \cite{chiralandvalley}.
Most interestingly, natural line defects occurring during growth, such as grain boundaries and topological defects, can also act as valley filters \cite{valleyline,valleyline2,valleyline3,valleyline4,valleylineexp1,valleylineexp2}.

Such valley separation can be seen to arise by perturbations belonging to different symmetry classes \cite{isotropic,symmetrybreaking,helen,elastic,scatteringtimes,biref,DavidGraphene}.
For example, adatom deposition that favors a sublattice (staggered field) leads to filtering of valley degenerate
currents \cite{symmetrybreaking,valleymassbarrier,valleymassbarrier2}.
Other methods include the deposition of graphene on pillar-superlattice arrays \cite{pillars,pillars2},
graphene decorated with Sn \cite{islands}, Au, Cu \cite{nanoparticles}, and other transition metals \cite{clustertransitionmetals},
or by graphene-azobenzene photo-controlled gated regions \cite{emanu}.
Although general symmetry characteristics of local scattering centers may give rise to valley selectivity and associated
valley Hall effect \cite{symmetrybreaking,ferreira2}, one wonders if such general conclusions can be obtained for
arrangements of symmetric scatterers that only collectively break lattice symmetries.

We address such question here. We study the scattering of Dirac fermions from a set of individually centrosymmetric scattering regions which, however, collectively break the spatial symmetries in graphene. We focus on the role played by mirror symmetry in these systems, the
constraints that its conservation imposes on impurity distribution, and the resulting scattering properties.
We show that impurity distributions that preserve mirror symmetry impose constraints on the scattering matrix that result in the absence of skew scattering.
In contrast, breaking mirror symmetry in a system allows for non-zero skew scattering;
having opposite sign for opposite valleys, this leads to the appearance of valley Hall effect in such system.
To quantify the impact of mirror symmetry, we consider pairs of arbitrarily oriented potential scatterers. We show that the skew parameter and the consequent
valley contribution to the Hall voltage depend on the size, strength, and location of the scatterers relative to the current direction, as
multiple scattering effects well beyond the diluted-impurity limit result in interesting effects \cite{cluster}. These findings suggest that the detection of a Hall
voltage in decorated graphene would interfere with other effects and should not be attributed solely to spin Hall effect \cite{colossal,nanoparticles}, as other
non-negligible contributions of symmetry breaking may contribute to the detected voltages \cite{symmetrybreaking,sroche}.

\textit{Multiple-center scattering of Dirac particles}. The scattering  from an arrangement of multiple perturbation (or impurity) centers in close proximity is described in graphene by the Hamiltonian
\begin{equation}\label{hamiltonian}
H=v_{F}(\alpha_{1}p_{x}+\alpha_{2}p_{y})+\sum_{l=1}^{N}{\hat{V}_{l}\Theta(R_{l}-|\vec{r} - \vec{r}_{l}|)}\, .
\end{equation}
This is written in the chiral basis $\psi=\left(\psi_{A,K},\psi_{B,K},\psi_{B,K'},\psi_{A,K'}\right)^{T}$ \cite{symmetrybreaking},
where $v_{F}$ is the Fermi velocity, $\alpha_{i}=\tau_{3}\otimes\sigma_{i}$ $(i=1,2,3)$, where $\sigma_{i}$ and
$\tau_{i}$ are Pauli matrices that act on the sublattice
index ($A,B$) and the valley index ($K,K'$), respectively.
It is also useful to define the identity $I=\sigma_{0}\otimes\tau_{0}$,  $\gamma^{5}=\tau_{3}\otimes\sigma_{0}$,   and
$\beta=\tau_{1}\otimes \sigma_{0}$, so that
$\hat{V}_{l}=V_{l}I$, where $V_{l}$ is the strength of the $l$-th potential scatterer.
Each of the $N$ scatterers in the cluster under consideration is assumed constant over a region of radius $R_l$, centered
at $\vec{r}_l$;  $\Theta$ is a Heaviside step function, and locations are measured with respect to a global coordinate
system, with $\vec{r}=(x,y)=(r,\theta)$. Away from the scattering region, the wave is assumed to recover its asymptotic form in unperturbed graphene.

\begin{figure}
  \centering
  \includegraphics[scale=0.35]{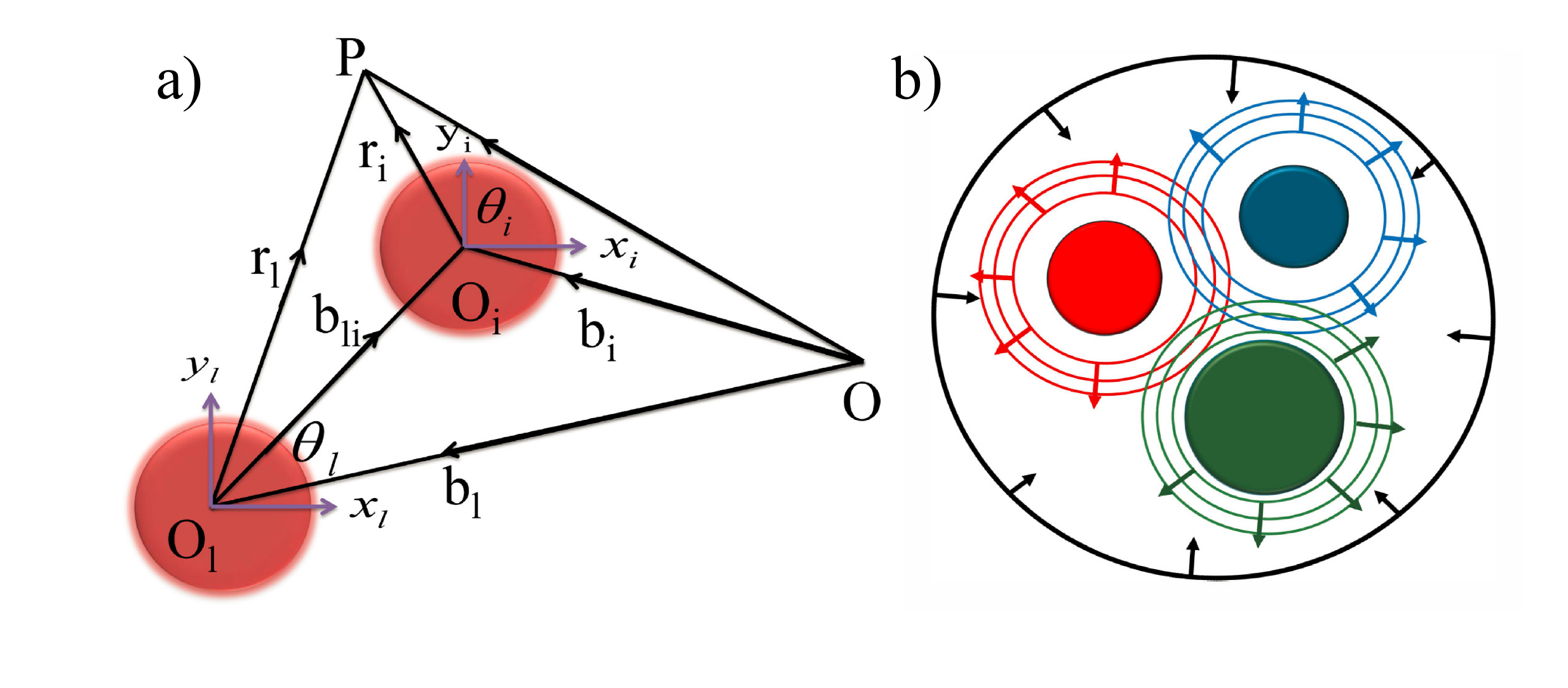}
  \caption{a) The global and local coordinate systems for two scattering centers. The local origins are defined from the center of the scatterers $i$ and $l$, while the global origin is $O$. b) Schematic of scattered and incoming waves in a system of multiple scatterers.}   \label{fig1}
\end{figure}

The continuity of the wave function in and out the different scattering centers allows the determination of the scattering matrix,
and through the far-field matching, one finds the scattering amplitudes of the problem. In order to take into account multiple
scattering events, the continuity of the wave function at all boundaries is enforced.  This is facilitated by
considering the local conservation of total angular momentum around each scatterer,
\begin{equation}
J_{z_{l}}=-i\partial_{\theta_{l}}+\frac{1}{2}\gamma^{5}\alpha_{3}\;,
\end{equation}
allowing us to label the local eigenstates by their angular momentum, such that $J_{z_{l}}\psi_{j}(r_{l},\theta_{l})=j\psi_{j}(r_{l},\theta_{l})$, where $j$ is a half-integer ($\hbar=1$).  This allows one to express the wave functions at any point in terms of
the local coordinates of a given scatterer via (Graf's)
addition theorems \cite{SuppInfo,Abramowitz}.
Within this formalism, as the perturbation vanishes asymptotically, the wave function is described by incoming and outgoing eigenstates of the free Hamiltonian, $H_{0}\psi=E\psi$, and can be written as
\begin{equation}\label{smult}
\psi =\psi^{(-)}_{j}+\sum_{j'}{S_{jj'}\psi^{(+)}_{j'}}\;,
\end{equation}
where $j$ and $j'$ are half-integers, $\psi^{(-)}_{j}$ is the incident front. The scattering matrix $S$ is in general not diagonal in angular momentum for a generic distribution of scattering centers.  However, it can be expressed in
terms of the local (angular momentum diagonal) matrices about each center, $s^l$
\begin{equation}\label{snm}
{S_{jj'}}=\sum_{l} {s^{l}_{j}J_{j'-j}(kb_{l})e^{i(j'-j)\varepsilon_{l0}} }\;,
\end{equation}
where  $b_{l}$ and $\varepsilon_{l0}$ are the distance and relative angle from the global origin of coordinates to
the center of the $l$-th impurity (scattering center), respectively, and $s^{l}_j$ is the (diagonal) scattering
matrix element describing the $l$-th center due to the $j$ partial wave.
These elements are determined by a set of $2N\times M$ linearly coupled equations
that consider $M=2|j_{\rm max}+\frac{1}{2}|$ angular momentum channels (see Supplement \cite{SuppInfo}).
The far-field scattering amplitude for a distribution of scatterers is
\begin{equation}\label{amplitude}
f(\theta)=\frac{e^{-i\pi/4}}{\sqrt{2\pi k}}\sum_{j}{(S_{j}-1)} e^{i(j-\frac{1}{2})\theta} \, ,
\end{equation}
with
$
S_{j}=\sum_{j'}{S_{jj'}}\;.
$

\textit{Symmetry considerations}. We now recall important symmetries of the low-energy Dirac Hamiltonian of graphene \cite{symmetrybreaking}.
The {\em time reversal} operation reverses the momentum and exchanges valley index in the eigenstate, so that
in the chiral basis this operator is defined as $\mathcal{T}=\beta\gamma^{5}\alpha_{1}\mathcal{C}$, where $\mathcal{C}$ is complex conjugation.
In two dimensions we have two orthogonal mirror axes, and can define corresponding parity (mirror) operators $\mathcal{P}_{x}$, and $\mathcal{P}_{y}$.
Their explicit definition depends on the underlaying orientation of the lattice; if one assumes that zigzag chains of graphene are along
the $x$-axis, then $\mathcal{P}_{x}$ requires changing of sublattice index and $(x,y)\rightarrow(x,-y)$, while keeping the valley untouched.
Meanwhile, $\mathcal{P}_{y}$ requires to change sublattice, valley, and $(x,y)\rightarrow(-x,y)$. These considerations lead one to write
$\mathcal{P}_{x}=\gamma^{5}\alpha_{1}$, with  $(x,y)\rightarrow(x,-y)$, and, $\mathcal{P}_{y}=\gamma^{5}\alpha_{1}\beta$, with  $(x,y)\rightarrow(-x,y)$.
The combination of the two reflections leads to inversion, equivalent here to a $\pi$-rotation of the lattice, $\mathcal{I}=\beta$, which exchanges sublattice, valley and $(x,y)\rightarrow (-x,-y)$ \cite{quntums,timerev}.

For the scattering problem, symmetries impose restrictions on the $S$ matrix and correspondingly on different observables, as we now
describe.
Assume an incoming circular wave approaching the scattering region containing a set of time reversal and parity invariant perturbations,
as described by Eq.\ (\ref{smult}). Considering  $P_{\pm}=(1\pm\gamma^{5})/2$ as the projectors into valley space ($+,-=K,K'$), such that, $
\mathcal{P}_{x}(P_{\pm}\psi)=\pm P_\pm \psi$, we have $\mathcal{P}_{x}\psi_{j}=i(-1)^{j+\frac{1}{2}}\gamma^{5}\psi_{-j}$ \cite{SuppInfo}.
Applying then the mirror symmetry operation to the state $\psi_{j}$ in Eq.~\ref{smult}, we get
\begin{equation}\label{parity1}
 \psi_{-j}=\psi^{(-)}_{-j}+\sum_{j'}{(-1)^{j'-j}\gamma^{5}\mathcal{P}_{x}S_{jj'}\mathcal{P}_{x}^{-1}\gamma^{5}\psi^{(+)}_{-j'}}\;.
\end{equation}
Comparing to the scattering of a circular wave of incident angular momentum $-j$,
\begin{equation}\label{smult2}
\psi_{-j}=\psi^{(-)}_{-j}+\sum_{j'}{S_{-jj'}\psi^{(+)}_{j'}}\;,
\end{equation}
one gets
\begin{equation}\label{parity2}
(-1)^{j'-j}\gamma^{5}\mathcal{P}_{x}S_{jj'}\mathcal{P}_{x}^{-1}\gamma^{5}=S_{-j-j'}\;.
\end{equation}

Similarly, we can use the time reversal operator $\mathcal{T}$ \cite{SuppInfo}, so that
\begin{equation}\label{t2}
(-1)^{j'-j}\gamma^{5}\mathcal{T}S_{jj'}\mathcal{T}^{-1}\gamma^{5}=S^{*}_{-j-j'}\;.
\end{equation}

Using the symmetries above results in constraints on the $S$ matrix elements,
$(-1)^{j'-j}S_{jj',\tau\tau}$$=$$S_{-j-j',\tau\tau}$$=$$S_{-j-j',\bar{\tau}\bar{\tau}}$ and $(-1)^{j'-j}S_{jj',\tau\bar{\tau}}$$=$$S_{-j-j',\tau\bar{\tau}}$$=-S_{-j-j',\bar{\tau}\tau}$, where $\tau=-\bar{\tau}$, and $\tau=\pm$ indicates the valley ($K,K'$) index. These in turn are reflected on the scattering amplitudes, such
that $f_{j,\tau\tau}$$=$$f_{-j,\tau\tau}$$=$$f_{-j,\bar{\tau}\bar{\tau}}$, and $f_{j,\tau\bar{\tau}}$$=$$f_{-j,\tau\bar{\tau}}$$=$$-f_{-j,\bar{\tau}\tau}$, where $f_{j}=S_j-1$, as given in Eq.\ \ref{amplitude}.

The asymmetry of the scattered waves about the incoming flux axis is quantified by the skew cross section,
\begin{eqnarray}
\sigma_{sk,\tau\tau'}&=& \int d \theta \; \sigma_{\tau,\tau'} (\theta) \, \sin \theta  \nonumber \\
				&=&\frac{1}{k}\sum_{j}{{\rm Im}(f_{j,\tau\tau'}f^{*}_{j+1,\tau\tau'})}\;,
\end{eqnarray}
where $\sigma_{\tau,\tau'} (\theta)$ is the scattering cross section valley matrix.\cite{SuppInfo}
In a system with both $\mathcal{P}_{x}$ and time reversal symmetry, we have
\begin{equation}
\sum_{j}{f_{j,\tau\tau'}f^{*}_{j+1,\tau\tau'}}=|f_{\frac{1}{2},\tau\tau'}|^{2}+\sum_{j\ge\frac{1}{2}}{2{\rm Re}(f_{j,\tau\tau'}f^{*}_{j+1,\tau\tau'})}\, ,
\end{equation}
from which it results that if both $\mathcal{P}_{x}$ and $\mathcal{T}$ are preserved by the perturbations, the system will not have skew scattering, and $\sigma_{sk,\tau\tau'}=0$.

If the perturbation potential of the multiple impurity assembly {\em lacks} $\mathcal{P}_{x}$ mirror symmetry, the conditions on the scattering
amplitudes are reduced to those imposed only
by time reversal, such that $f_{j,\tau\tau}=f_{-j,\bar{\tau}\bar{\tau}}$, and $f_{j,\tau\bar{\tau}}=-f_{-j,\bar{\tau}\tau}$.  This leads to
\begin{eqnarray}
&&\sigma_{sk,\tau\bar{\tau}}=0\;, \nonumber \\
&&\sigma_{sk,\tau\tau}=-\sigma_{sk,\bar{\tau}\bar{\tau}}\;,
\end{eqnarray}
which would give rise to splitting of valley currents in real space. This is associated with the appearance of  transverse valley currents, $j_{VH}$, and
characterized by the valley Hall angle \cite{symmetrybreaking}, $\tan \Theta_{VH} \simeq \Theta_{VH}=j_{VH}/j_{inc}$, in terms of the incident current $j_{inc}$~\cite{SuppInfo}.
At zero temperature and in the absence of the side jump effect \cite{sinova}, this angle is given by the valley skew parameter
\begin{equation}
\gamma_{V}=(\gamma_{K}-\gamma_{K'})/2 \, ,
\end{equation}
where $\gamma_{\tau}=\sum_{\tau'}{\sigma_{sk,\tau'\tau}/\sigma_{tr,\tau'\tau}}$, and $\sigma_{tr}$ is the transport cross section \cite{SuppInfo}.
It is clear that whenever  $\mathcal{P}_{x}$ is conserved by the perturbation, $\gamma_{V}=0$; in contrast, when $\mathcal{P}_{x}$ is broken,
$\gamma_{V}\ne0$ is allowed, which would indicate a non-vanishing valley Hall effect. Here we emphasize that the spatial distributions of impurities that lead to the appearance of a valley Hall strongly depend on the nature of the perturbations enhanced by these impurities, including intervalley scattering, as shown in the supplementary material~\cite{SuppInfo}.

\begin{figure}
  \centering
  \includegraphics[scale=0.55]{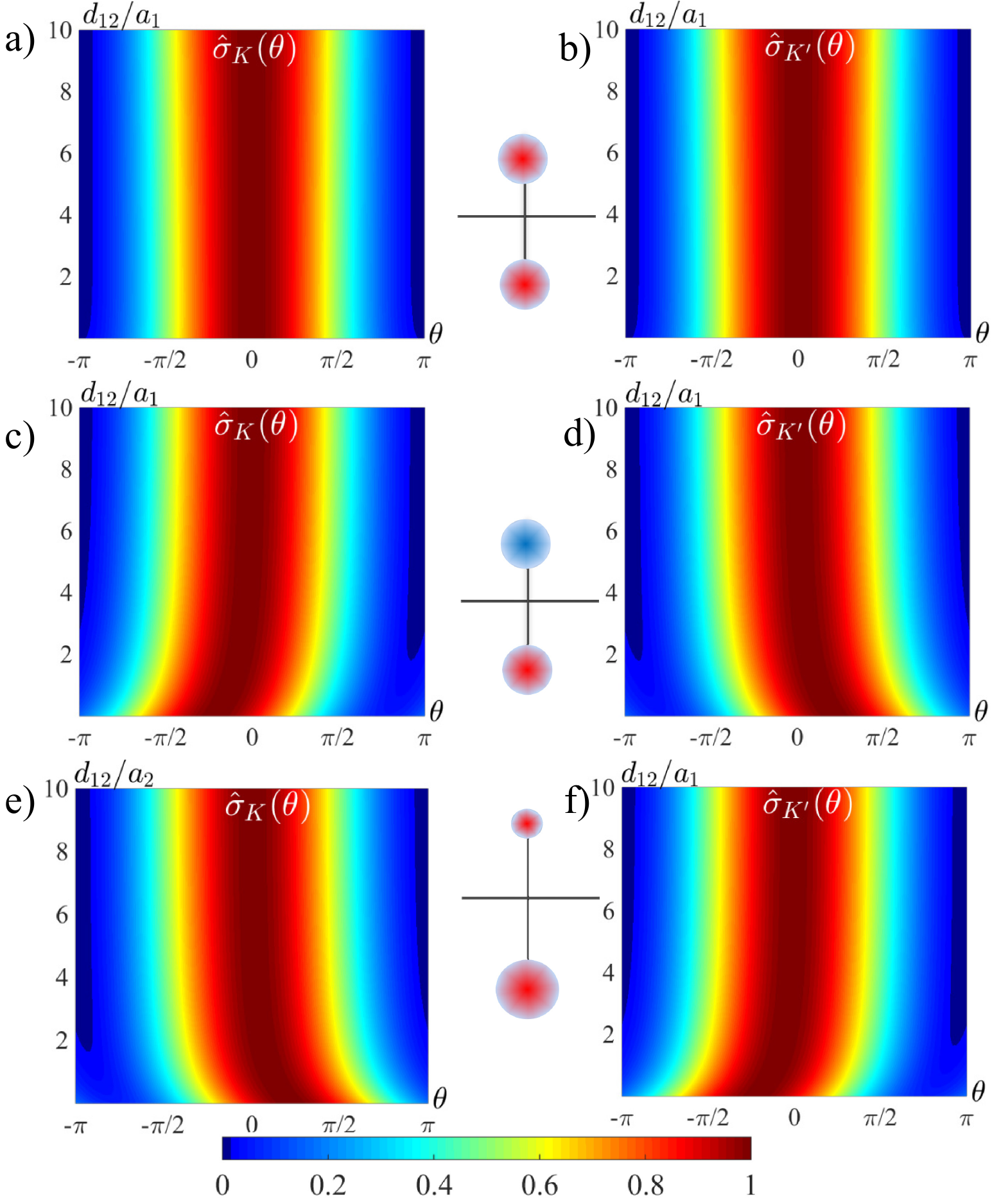}
  \caption{Differential cross section maps, $\hat{\sigma}_{\tau}(\theta)=\sigma_{\tau\tau}(\theta)/{\rm max}(\sigma_{\tau\tau}(\theta))$, where $\tau=(K,K')$
for incoming electrons along the $\hat{x}$  direction with energy $E=1$ meV and different configurations of scattering potentials.
a) \& b) Mirror symmetric configuration for two identical scattering centers of radius $a_{1}$$=$$a_{2}$$=$$10$ {\rm \AA}, and strength $V_{1}$$=$$V_{2}$$=$$2$ eV.\@
The two centers have a separation $r_{12}$$=$$d_{12}$$+$$2a$. Notice that as $\mathcal{P}_{x}$ is preserved, the differential cross sections for the two valleys are identical--panel a) $\sigma_{K}$, and b) $\sigma_{K'}$.
c) \& d) Show $\mathcal{P}_{x}$ asymmetric configuration with $V_{1}$=$-V_{2}$$=$$2$ eV, and $a_{1}$$=$$a_{2}$$=$$10$ {\rm \AA}.
Notice that $\hat{\sigma}_{K}(\theta)\ne\hat{\sigma}_{K'}(\theta)$ for distances $d_{12} \lesssim 4a_1$, where multiple scattering effects are pronounced.
In e) \& f) $\mathcal{P}_{x}$ asymmetric configuration with $2a_{1}=a_{2}=10$\AA, and $V_1$$=$$V_{2}$$=$$2$ eV.\@
Effects of asymmetry at short separations are also evident.}  \label{fig2}
\end{figure}

\textit{Representative system}. To exemplify the effects of mirror symmetry breaking, we consider a minimal system of two centrosymmetric impurities.
As such, each scattering center respects all symmetries of the Hamiltonian and does not induce valley splitting or mixing. However, as we will
show, such minimal set may break parity and result in skew scattering and valley splitting
for different arrangements of these two impurities. We emphasize that if mirror symmetry $\mathcal{P}_{x}$ is
maintained, then the $K$ and $K'$ valley components of the scattered waves follow identical trajectories with identical differential cross sections and yield
no valley Hall effect. This is the case when the impurities are identical in size, potential strength, {\em and} aligned along the $x$ or $y$ axes,
as shown in Fig.\ \ref{fig2}a and b (also see [\onlinecite{SuppInfo}]). The scenario drastically changes, however, if the impurities have different size and/or strength.
For example, when the two impurities have different potential strength but equal size and are aligned along the $y$ axis, Fig.\ \ref{fig2}c \& d, $\mathcal{P}_{x}$ mirror symmetry is broken.  In that case, the differential cross sections for the $K$ and $K'$ sectors are remarkably different at
short separation, where multiple scattering effects are more important; the typical single-scatterer differential cross section is recovered at large
separations, as one would expect \cite{isotropic}. A similar effect can be seen in Fig.~\ref{fig2}e \& f, when the size of the two impurities is
different. In these cases, the contrast in differential cross sections results in the separation of the $K$ and $K'$ scattered electrons in space--especially when
the perturbation centers are close to each other ($d_{12} \lesssim 2a_2$ for these parameters).

\begin{figure}
  \centering
  \includegraphics[width=0.5\textwidth]{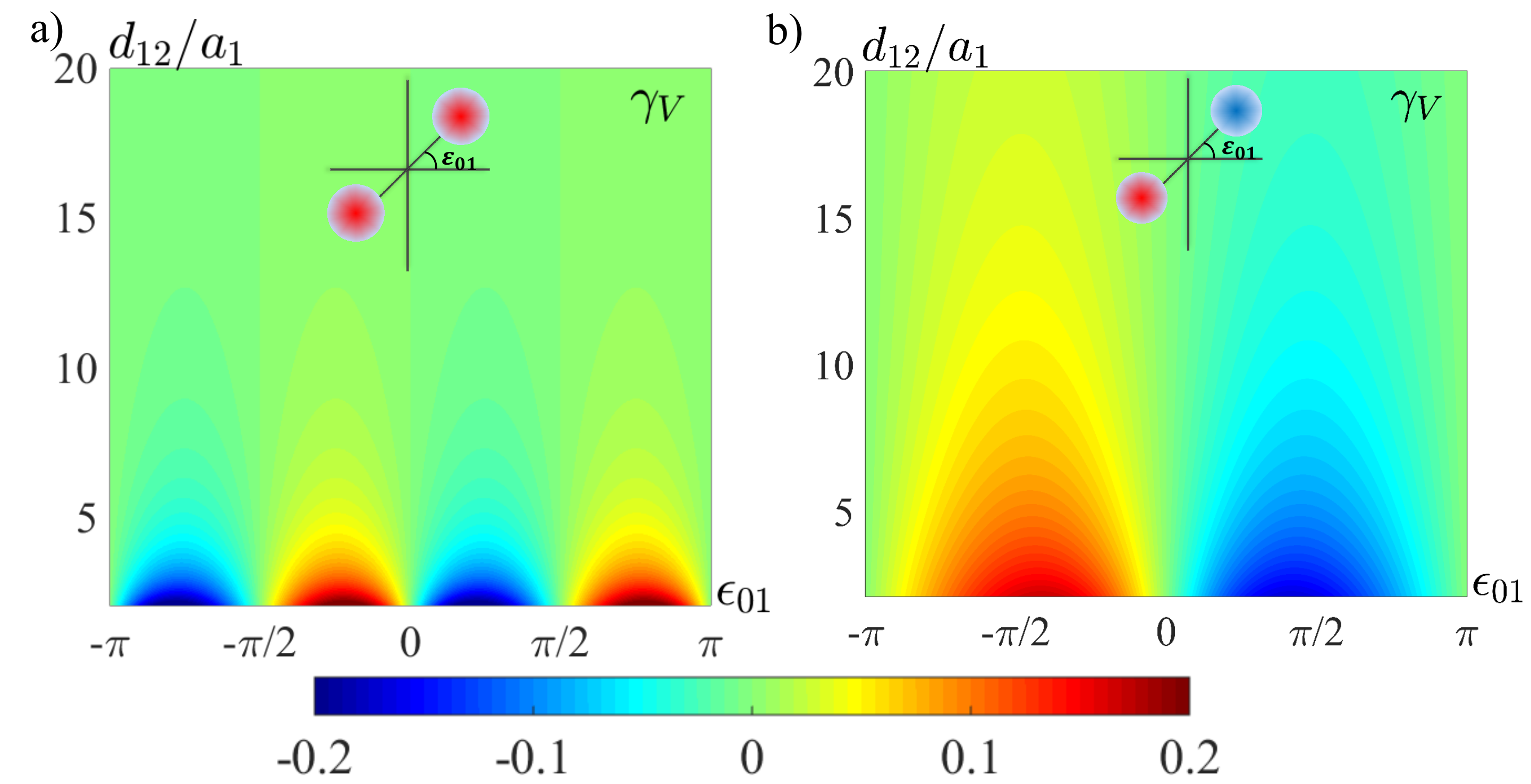}
  \caption{Valley skew parameter, $\gamma_{V}$ as a function of the spacing between potential impurities, and angle $\epsilon_{01}$. Notice that $\epsilon_{01}$ is the relative angle from impurity $1$ to the global origin, and we fix $\epsilon_{02}=\pi+\epsilon_{01}$. a) $V_{1}$$=$$V_{2}$$=$$2$eV and $a_{1}$$=$ $a_{2}$$=$$10\;{\rm \AA}$. Notice that $\gamma_{V}=0$ for $\epsilon_{01}=0,\pm \pi/2, \pm \pi$ regardless of the spacing $d_{12}$, since $\mathcal{P}_{x}$ is preserved for those configurations. b) $V_{1}$$=$$-V_{2}$$=$$2$ eV, $a_{1}$$=$$a_{2}$$=$$10\; \AA$. Here, notice that $\gamma_{V}=0$ only for $\epsilon_{01}=0$, since $\mathcal{P}_{x}$ is conserved only for that configuration. }\label{fig3}
\end{figure}

Breaking of $\mathcal{P}_{x}$ mirror symmetry is directly reflected in the appearance of skew scattering.
To show how, we change the relative orientation of the two impurities with respect
to the incident plane wave direction. As seen in Fig.\ \ref{fig3}a for a system with high symmetry, when both impurities have the same size and strength, $\mathcal{P}_{x}$ symmetry is recovered for two configurations, $\epsilon_{01}$$=$$(0, \pm\pi)$, and $\pm\pi/2$; in these configurations the skew
parameter clearly vanishes, $\gamma_{V}$$=0$. Away from these, the skew parameter alternates in sign and reaches a maximum amplitude at
$\epsilon_{01}=\pm\pi/4,\pm3\pi/4$, decaying to zero for $d_{12} \gtrsim 5a_1$.

For a system with lower symmetry, where the two impurity potentials are different, the skew parameter shows different behavior, as displayed in
Fig.\ \ref{fig3}b.  As the mirror symmetry is recovered for configurations with $\epsilon_{01}=0,\pm \pi$,  the skewness is seen to vanish.  The skew parameter
$\gamma_{V}$ alternates in sign away from these $\epsilon_{01}$ values and has a slower decay with $d_{12}$, remaining non-zero for $d_{12} \lesssim 15a_1$.

Note that the symmetry arguments hold for any number of impurities that produce time reversal invariant perturbations \cite{SuppInfo}, which suggests that for a generic cluster of impurities
that break mirror symmetry, the skewness parameter and valley Hall angle will be nonzero, leading in general to the appearance of valley Hall effect.

\emph{Discussion}.  Based on these results and the symmetry analysis presented, we now consider some recent experimental results on graphene
that alter the deposition of adatoms \cite{colossal,nanoparticles}.  These experiments have shown sizeable nonlocal resistance in
decorated samples, and interpreted such as due to the appearance of spin Hall effect via skew scattering.  The nature
of the Hall effect in these systems depends on the proximity and distribution of the adatoms, as our analysis shows.  In the dilute limit, where multiple scattering effects are
negligible, one would not expect to see valley Hall effect induced by skew scattering from symmetric and/or point impurities. In that case, a non-vanishing
nonlocal resistance may be seen as an indication of spin Hall effect. However, it has been shown that even in the dilute limit,\cite{symmetrybreaking}  if the
adatoms break spatial symmetries and enhance spin-orbit interactions, the non-local resistance has both contributions from the spin skewness but also from
valley skewness. This would lead to an overestimate of the spin-orbit enhancement in the system.  Our analysis here, together with
[\onlinecite{symmetrybreaking}], suggest that non-local resistance measurements in systems of decorated graphene are not sufficient to determine the nature of
the Hall effect or to estimate the spin-orbit coupling in the system. This, incidentally, is in agreement with recent theoretical results \cite{sroche}, and
experiments \cite{noSH1,noSHE2}.

It should be noted that in systems where the distance between scattering centers is large (larger than the dephasing length, $L_\phi$), the electronic scattering from one impurity to the next are essentially independent. Additionally, the total skewness of the system may be reduced, even if mirror symmetry is broken by each cluster, as the skewness gained by scattering from one cluster can be inverted by the next. This averaging of the skew parameter due to randomness in cluster irregularities, suggests that skew scattering and valley filtering would be better probed in experiments with a high degree of control over impurity potential configurations \cite{emanu,pillars,pillars2,emabook}.
​In fact, designing properly asymmetric clusters well contained within a dephasing length ($a_i,d_{12} \ll L_\phi \simeq 1 \mu$m), but similarly oriented on the sample would result in sizable valley Hall filters.  Such filters could be built by lithographically controlled deposition of clusters on the graphene surface and provide an interesting tool in valleytronics.

\emph{Conclusions.} We have studied the effects of impurity distribution on the scattering of Dirac fermions in graphene. We have shown the
importance of constraints that symmetries impose on the scattering matrix.  For a set of potential impurities that break mirror symmetry
with respect to the axis of scattering, the differential cross sections for different valleys become inequivalent, and a nonzero Hall angle is expected.
This broken symmetry results in the appearance of valley Hall effect. Our results and discussion also suggest that a set of impurities with well tailored features would exhibit different degrees of skewness which could be probed and exploited to produce valley polarized currents
at will in graphene systems.

\emph{Acknowledgments.} We acknowledge support from NSF grants DMR-1508325 (Ohio) and DMR-1410741 and DMR-1151717 (LSU).

  \bibliography{references}
\end{document}